\documentclass{ws-ijmpd}

\begin{document}

\markboth{B. Blossier {\em et al.} [Alpha Collaboration]}
{Nonperturbative HQET at Order $1/m$}

%
\catchline{}{}{}{}{}
%

\title{Nonperturbative HQET at Order $1/m$}

\author{Beno\^it Blossier\\
        Laboratoire de Physique Th\'eorique, B\^atiment 210,
        Universit\'e Paris XI, \\F-91405 Orsay Cedex, France}
\author{Georg von Hippel\\
        Institut f\"ur Kernphysik,
        University of Mainz, D-55099 Mainz, Germany}
\author{Nicolas Garron\\
        School of Physics and Astronomy, University of Edinburgh, 
        Edinburgh EH9 3JZ, UK}
\author{Tereza Mendes\thanks{Speaker.}\\
        IFSC, University of S\~ao Paulo,
        C.P. 369, CEP 13560-970, S\~ao Carlos SP, Brazil}

\maketitle

\begin{history}
\received{Day Month Year}
\revised{Day Month Year}
\comby{Managing Editor}
\end{history}

\begin{abstract}
  We summarize first results for masses and decay constants
  of bottom-strange (pseudo-scalar and vector) mesons from
  nonperturbatively renormalized heavy-quark effective theory (HQET), 
  using lattice-QCD simulations in the quenched approximation.
\end{abstract}

\keywords{B physics; lattice QCD; heavy-quark effective theory.}

\section{Introduction}

The study of B physics is essential to determine the flavor
structure of the Standard Model, through knowledge of the 
Cabibbo-Kobayashi-Maskawa (CKM) matrix describing quark mixing 
and CP violation, which may be associated with the lack of 
symmetry between matter and anti-matter in the Universe. 
In fact, since the amount of baryons in the Universe predicted using 
the CKM mechanism is several orders of magnitude smaller
than what is observed by astronomers, extensions of the 
Standard Model propose additional sources of CP violation,  
which must be tested against Standard-Model predictions.
B mesons provide the ideal environment for such tests.\cite{Antonelli:2009ws}
In particular, high-precision theoretical inputs are needed
for hadronic matrix elements, which may be computed starting
from the gauge theory itself using numerical simulations of lattice QCD.
At present, however, it is not yet feasible to perform simulations 
on lattices that can simultaneously represent the two relevant scales 
of B physics:
the low energy scale $\Lambda_{\rm QCD}$, requiring
large physical lattice size, and the high energy scale of
the b-quark mass $m_b$, requiring very small lattice spacing $a$.
An approximate framework is therefore needed, but one should strive
to achieve sufficiently precise results, otherwise the task of
overconstraining the parameters of the Standard Model is compromised.

A promising such framework is to consider (lattice) heavy-quark effective 
theory (HQET), which allows for an elegant theoretical treatment, with 
the possibility of fully nonperturbative 
renormalization.\cite{Heitger:2003nj,Sommer:2006sj}
The approach is briefly described as follows.
HQET provides a valid low-momentum description for systems with one
heavy quark, with manifest heavy-quark symmetry in the limit 
$m_b \to \infty$. The heavy-quark flavor 
and spin symmetries are broken at finite values of $m_b$
respectively by kinetic and spin terms, with first-order corrections 
to the static Lagrangian parametrized by $\omega_{\rm kin}$ and 
$\omega_{\rm spin}$
\begin{equation}
{\cal L}^{\rm HQET}
\;=\; \overline{\psi}_{\!h}(x)\,D_0\,\psi_h(x)
\,-\, \omega_{\rm kin}\, {\cal O}_{\rm kin} 
\,-\, \omega_{\rm spin}\,{\cal O}_{\rm spin}\,,
\end{equation}
where 
\begin{equation}
{\cal O}_{\rm kin}\;=\; \overline{\psi}_{\!h}(x)\,{\bf D}^2\,\psi_h(x)\,,
\quad\quad
{\cal O}_{\rm spin}\;=\; 
\overline{\psi}_{\!h}(x)\,{\bf \sigma}\cdot{\bf B}\,\psi_h(x)\,.
\end{equation}
These ${\rm O}(1/m_b)$ corrections are incorporated by an expansion 
of the statistical weight in $1/m_b$
such that ${\cal O}_{\rm kin}$, ${\cal O}_{\rm spin}$ are
treated as insertions into static correlation functions.
This guarantees the existence of a continuum limit, with results that
are independent of the regularization, provided that the renormalization
be done nonperturbatively. 

As a consequence, expansions for masses and decay constants 
are given respectively by
\begin{equation}
m_B \;=\; m_{\rm bare} \,+\,E^{\rm stat}
\,+\, \omega_{\rm kin} \, E^{\rm kin} \,+\,
\omega_{\rm spin} \, E^{\rm spin}
\end{equation}
and
\begin{equation}
f_B\,\sqrt{\frac{m_B}{2}} \;=\; Z_A^{\rm HQET}\,p^{\rm stat}\,
(1\,+\, c_A^{\rm HQET} \, p^{\delta A}
\,+\, \omega_{\rm kin} \, p^{\rm kin} \,+\,
\omega_{\rm spin} \, p^{\rm spin}) \,,
\end{equation}
where the parameters $m_{\rm bare}$ and $Z_A^{\rm HQET}$ are written as
sums of a static and an ${\rm O}(1/m_b)$ term (denoted respectively with the 
superscripts ``${\rm stat}$'' and ``$1/m_b$'' below), and 
$c_A^{\rm HQET}$ is of order $1/m_b$.
Bare energies ($E^{\rm stat}$, etc.) and matrix elements 
($p^{\rm stat}$, etc.) are computed in the numerical 
simulation.

The divergences (with inverse powers of $a$) in the above parameters 
are cancelled through the nonperturbative renormalization, 
which is based on a matching of HQET parameters to QCD on lattices of 
small physical volume 
--- where fine lattice spacings can be considered --- and extrapolation 
to a large volume by the step-scaling method.
Such an analysis has been recently completed for the quenched 
case.\cite{Blossier:2010jk}
In particular, there are nonperturbative (quenched) 
determinations of the static coefficients $m_{\rm bare}^{\rm stat}$
and $Z_{A}^{\rm stat}$ for HYP1 and HYP2 static-quark 
actions\cite{DellaMorte:2003mn} at the physical b-quark mass, and similarly
for the ${\rm O}(1/m_b)$ parameters
$\omega_{\rm kin}$, $\omega_{\rm spin}$,
$\,m_{\rm bare}^{1/m_b}$, $\,Z_{A}^{1/m_b}$ and $c_A^{\rm HQET}$.

The newly determined HQET parameters are very precise (with errors of 
a couple of a percent in the static case) and show the expected behavior 
with $a$. They are used in our calculations reported here, to perform
the nonperturbative renormalization of the (bare) observables computed
in the simulation. Of course, in order to keep a high precision, also 
these bare quantities have to be accurately determined. This is accomplished
by an efficient use of the generalized eigenvalue problem
(GEVP) for extracting energy levels $E_n$ and matrix elements, as
described below.


A significant source of systematic errors in the determination of
energy levels in lattice simulations is the contamination from
excited states in the time correlators 
\begin{equation}
C(t) \;=\; \,\langle O(t)\,O(0)\rangle\,  \;=\;
\sum_{n=1}^{\infty}\, |\langle n|\,\hat{O}\,|0 \rangle |^2\;e^{-E_n t}
\end{equation}
of fields $\,O(t)$ with the quantum numbers of a given bound state. 

Instead of starting from simple local fields $O$
and getting the (ground-state) energy from an effective-mass plateau in 
$C(t)$ as defined above, it is then advantageous to consider all-to-all 
propagators\cite{Foley:2005ac} and to solve, instead, the GEVP 
\begin{equation}
C(t)\,v_n(t,t_0)\;=\;
\lambda_n(t,t_0)\,C(t_0)\,v_n(t,t_0)\,,
\end{equation}
where $t>t_0$ and $C(t)$ is now a matrix of correlators, given by
\begin{equation}
C_{ij}(t)\,\;=\;\,\langle O_i(t) O_j(0)\rangle
\,\;=\;\, \sum_{n=1}^\infty {\rm e}^{-E_n t}\, \Psi_{ni}\Psi_{nj}\,,\quad
i, j = 1,\ldots, N\,.
\end{equation}
The chosen interpolators $O_i$ are taken (hopefully) 
linearly independent, e.g.\ they may be built from the 
smeared quark fields using $N$ different smearing levels. 
The matrix elements $\Psi_{ni}$ are defined by
\begin{equation}
\Psi_{ni} \;\equiv\; (\Psi_n)_i \;=\; \langle n|\hat O_i|0\rangle\;,
\quad\; \langle m|n\rangle \,=\,\delta_{mn}\,.
\end{equation}

One thus computes $C_{ij}$ for the interpolator basis $O_i$ from 
the numerical simulation, then gets effective energy levels $E_n^{\rm eff}$ 
and estimates for the matrix elements $\Psi_{ni}$ from the
solution $\lambda_n(t,t_0)$ of the GEVP at large $t$. For the energies
\begin{equation}
E_n^{\rm eff}(t,t_0)\;\equiv\; 
{1\over a} \, \log{\lambda_n(t,t_0) \over \lambda_n(t+a,t_0)}
\end{equation}
it is shown\cite{Luscher:1990ck}
that $E_n^{\rm eff}(t,t_0)$ converges exponentially as $t\to\infty$ 
(and fixed $t_0$) to the true energy $E_n$. However, since the
exponential falloff of higher contributions may be slow,
it is also essential to study the convergence as a function of
$t_0$ in order to achieve the required efficiency for the method.
This has been recently done,\cite{Blossier:2009kd} by explicit application 
of (ordinary) perturbation theory to a hypothetical truncated problem 
where only $N$ levels contribute. The solution in this case is exactly 
given by the true energies, and corrections due to the higher states 
are treated perturbatively. We get
\newcommand{\aeff}{\hat {\cal A}_n^{\rm eff}}
\newcommand{\qeff}{\hat {\cal Q}_n^{\rm eff}}
\newcommand{\corren}{\varepsilon_{n}}
\newcommand{\corrpn}{\pi_{n}}
\begin{equation}
E_n^{\rm eff}(t,t_0) \;=\; 
E_n \,+\, {\varepsilon_{n}(t,t_0)}\,
\end{equation}
for the energies and
\begin{equation}
{\rm e}^{-\hat H t}(\qeff(t,t_0))^\dagger|0\rangle \;=\;
|n\rangle \,+\, \sum_{n'=1}^\infty  \pi_{nn'}(t,t_0)
\, |n'\rangle
\end{equation}
for the eigenstates of the Hamiltonian, which may be estimated through
\begin{eqnarray}
    \qeff(t,t_0) &=& R_n \,(\hat O\,,\,v_n(t,t_0)\,) \,, \\[2mm]
R_n &=& {\left(v_n(t,t_0)\,,\, C(t)\,v_n(t,t_0)\right)}^{-1/2}
\; \left[{\lambda_n(t_0+a,t_0) \over \lambda_n(t_0+2a,t_0)}\right]^{t/2}\,.
\end{eqnarray}

In our analysis we see that, due to cancellations of $t$-independent 
terms in the effective energy, the first-order corrections in 
${\varepsilon_{n}(t,t_0)}$ are independent of $t_0$ and very strongly 
suppressed at large $t$. We identify two regimes:
1) for $t_0 \,<\, t/2$, the 2nd-order corrections dominate and
their exponential suppression is given by the smallest energy gap 
$\,|E_m-E_n|\equiv \Delta E_{m,n}\,$ between level $n$ and its neighboring 
levels $m$; and 2) for $t_0 \,\geq\, t/2$,
the 1st-order corrections dominate and the suppression is given by the 
large gap $\Delta E_{N+1,n}$. 
Amplitudes $\,\pi_{nn'}(t,t_0)\,$ get main contributions
from the first-order corrections. For fixed $t-t_0$ these are also 
suppressed with $\Delta E_{N+1,n}$.
Clearly, the appearance of large energy gaps in the second regime
improves convergence significantly. We therefore work with $t$, $t_0$
combinations in this regime.


\def\first{1/m_b}
\def\stat{\rm {stat}}

A very important step of our approach is to realize that the same 
perturbative analysis may be applied to get the $1/m_b$ 
corrections in the HQET correlation functions mentioned previously
\begin{equation}
  C_{ij}(t) \;=\; C_{ij}^{\stat}(t) \,+\,
  \omega \,C_{ij}^{\first}(t) \,+\, {\rm O}(\omega^2)\,,
\end{equation}
where the combined ${\rm O}(1/m_b)$ corrections are symbolized by 
the expansion parameter $\omega$.
Following the same procedure as above, we get similar exponential
suppressions (with the static energy gaps) for static and ${\rm O}(1/m_b)$ 
terms in the effective theory. We arrive at
\begin{equation}
    E_n^{\rm eff}(t,t_0) \;=\; E_n^{{\rm eff},{\rm stat}}(t,t_0)
     +\omega E_n^{{\rm eff},{1/m_b}}(t,t_0) +{\rm O}(\omega^2)
\end{equation}
with
\begin{eqnarray}
    E_n^{{\rm eff},{\rm stat}}(t,t_0) &=&
    E_n^{\stat} \,+\, \beta_n^{\stat} \,
    {\rm e}^{-\Delta E_{N+1,n}^{\stat}\, t}+\ldots\,, 
\label{effenergies}
\\[2mm]
    E_n^{\rm eff,\first}(t,t_0) &=&
    E_n^{\first} \,+\, [\,\beta_n^{\first}
    \,-\, \beta_n^{\stat}\,t\;\Delta E_{N+1,n}^{\first}\,]
    {\rm e}^{-\Delta E_{N+1,n}^{\stat}\, t}+\ldots\, .
\end{eqnarray}
and similarly for matrix elements.
Preliminary results of our application of the methods described 
in this section were presented recently\cite{Blossier:2009mg}
and are summarized in the next section. A more detailed version of this 
study will be presented elsewhere.\cite{inprep}


\section{Results}

We carried out a study of static-light B$_{\rm s}$-mesons in 
quenched HQET with the nonperturbative parameters described in the
previous section,
employing the HYP1 and HYP2 lattice actions for the static quark
and an ${\rm O}(a)$-improved Wilson action for the strange quark
in the simulations. The lattices considered were of the form
$L^3 \times 2L$ with periodic boundary conditions. We took
$L\approx1.5$ fm and lattice spacings $0.1$ fm, $0.07$ fm 
and $0.05$ fm, corresponding respectively to
$\beta=6.0219$, $6.2885$ and $6.4956$.
We used all-to-all strange-quark propagators constructed from
approximate low modes, with 100 configurations.  
Gauge links in interpolating fields were smeared with 3 iterations 
of (spatial) APE smearing, whereas Gaussian smearing (8 levels) was used
for the strange-quark field.
A simple $\gamma_0\gamma_5$ structure in Dirac space was taken
for all 8 interpolating fields. 
Also, the local field (no smearing) was included in order
to compute the decay constant.


The resulting ($8\times8$) correlation matrix may be
conveniently truncated to an $N\times N$ one and the GEVP
solved for each $N$, so that results can be studied as a function of $N$.
We then pick a basis from unprojected interpolators, by
sampling the different smearing levels (from 1 to 7) as
$\{1,7\}$,$\{1,4,7\}$, etc. We perform fits of the various energy 
levels and values of $N$ to the behavior in Eq.\ (\ref{effenergies})
and extract our results from the predicted plateaus.
Next, we take the continuum limit, extrapolating our results to 
$a\to 0$. We see that the correction to the ground-state energy
due to terms of order $1/m_b$, which is positive for finite $a$,
is quite small (consistent with zero) in the continuum limit.
Our results for the pseudoscalar meson 
decay constant, both in the static limit and including ${\rm O}(1/m_b)$
corrections, are shown in terms of the combination 
$\Phi^{\rm HQET}\equiv F_{\rm PS}\,\sqrt{m_{\rm PS}}/C_{\rm PS}$, 
where $C_{\rm PS}(M/\Lambda_{\rm QCD})$ 
is a known matching function and $\Phi^{\rm RGI}$ denotes the
renormalization-group-invariant matrix element of the static 
axial current.\cite{DellaMorte:2007ij}
These two continuum extrapolations are shown in comparison with
fully relativistic heavy-light (around charm-strange) 
data\cite{DellaMorte:2007ij} in Fig.\ \ref{fig:fBs_comp} below.
Note that, up to perturbative corrections of order $\alpha^3$
in $C_{\rm PS}$, HQET predicts a behavior $const. + {\rm O}(1/r_0 m_{\rm PS})$
in this graph. Surprisingly no $1/(r_0 m_{\rm PS})^2$ terms are 
visible, even with our rather small errors.


\begin{figure}
\ \vspace*{-6mm}
\begin{center}
\includegraphics[width=10truecm]{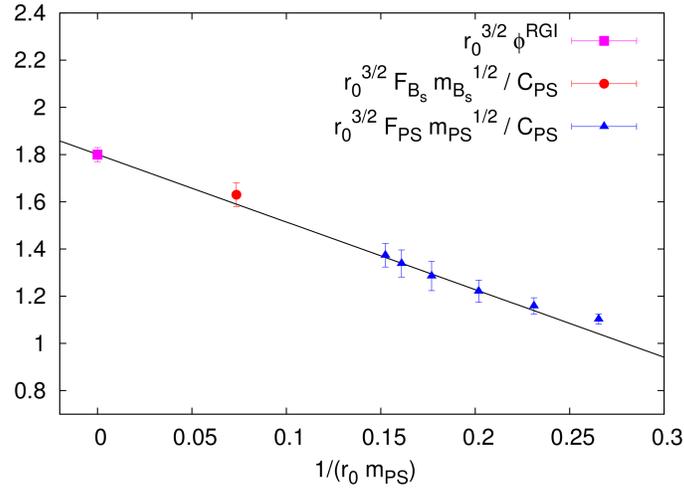}
\ \vspace*{-3mm}
\caption{Comparison of the continuum values for the pseudoscalar
meson decay constant from Fig.\ 4
to fully relativistic data in the charm region. The solid line is a
linear interpolation between the static limit and the points around
the charm-quark mass, which corresponds to $\,1/r_0\, m_{\rm PS}\approx 0.2$.
}
\label{fig:fBs_comp}
\end{center}
\ \vspace*{-6mm}
\end{figure}


\section{Conclusions}

The combined use of nonperturbatively determined HQET parameters 
(in action and currents) and efficient GEVP allows us to reach 
precisions of a few percent in matrix elements and of a few MeV 
in energy levels, even with only a moderate number of configurations. 
The method is robust with respect to the choice of interpolator basis.
All parameters have been determined nonperturbatively and in 
particular power divergences are completely subtracted.
We see that HQET plus ${\rm O}(1/m_b)$ corrections at the b-quark 
mass agrees well with an interpolation between the static point and the
charm region, indicating that linearity in $1/m$ extends even to 
the charm point. A corresponding study for $N_f=2$ is in progress.


\vskip 3mm
\noindent
{\bf Acknowledgements.}
This work is supported by the  DFG
in the SFB/TR 09, and by the EU Contract 
No.\ MRTN-CT-2006-035482, ``FLAVIAnet''. 
T.M. thanks the A. von Humboldt Foundation;
N.G. thanks the MICINN grant FPA2006-05807, the
Comunidad Aut\'onoma de Madrid programme HEPHACOS P-ESP-00346 and
the Consolider-Ingenio 2010 CPAN (CSD2007-00042).


\end{document}